# Institutional Repository *saber.ula.ve*: A testimonial perspective

Y. Briceño, H.Y. Contreras, L. A. Núñez, F. Salager-Meyer, A. Rojas, R. Torréns

*Abstract*

In this paper, we describe our decade-long experience of building and operating one of the most active Institutional Repository in the world: www.saber.ula.ve <http://www.saber.ula.ve> (University of the Andes, Mérida-Venezuela). In order to share our experience with other institutions, we firstly explain the steps we followed to preserve and disseminate the scientific production of the University of Los Andes' researchers. We then present some recent quantitative results about our repository activities and we outline some methodological guidelines that could be applied in order to replicate similar experiences.

These guidelines list the ingredients or building blocks as well as the processes followed for developing and maintaining the services of an Institutional Repository. These include technological infrastructure; institutional policies on preservation, publication and dissemination of knowledge; recommendations on incentives for open access publication; the process of selection, testing and adaptation of technological tools; the planning and organization of services, and the dissemination and support within the scientific community that will eventually lead to the adoption of the ideas that lie behind the open access movement.

We summarize the results obtained regarding the acceptance, adoption and use of the technological tools used for the publication of our institution's intellectual production, and we present the main obstacles encountered on the way.

Y. Briceño belongs to the Centro Nacional de Cálculo Científico, Universidad de Los Andes, Parque Tecnológico de Mérida, Mérida 5101, Venezuela;; e-mail: ysabelbr@ula.ve.

H.Y. Contreras belongs to the Centro de Simulación y Modelos, Departamento de Computación, fac. Ingeniería, Universidad de Los Andes, Mérida 5101, Venezuela; e-mail: hyelitza@ula.ve.,

L. A. Núñez belongs to the Centro de Física Fundamental, Dept. Física, Facultad de Ciencias, Universidad de Los Andes, Mérida 5101, Venezuela; e-mail: nunez@ula.ve.

F. Salager-Meyer coordinates the Grupo de Investigación Multidisciplinario del Análisis del Discurso Científico, Facultad de Medicina, Universidad de Los Andes, Mérida 5101, Venezuela; e-mail: francoise.sm@gmail.com

A. Rojas belongs to the Centro Nacional de Cálculo Científico, Universidad de Los Andes, Parque Tecnológico de Mérida, Mérida 5101, Venezuela; e-mail: ascanio@ula.ve.

R. Torréns belongs to the Centro Nacional de Cálculo Científico, Universidad de Los Andes, Parque Tecnológico de Mérida, Mérida 5101, Venezuela; e-mail: torrens@ula.ve.

I. INTRODUCTION

The progress and current availability of Information and Communication Technology (ICT) are causing profound changes in scholarly communication by creating new practices associated with the visibility and flow of information from academic research centers. Broadly speaking, these new practices have been made possible thanks to the unrestricted access to scientific information with new ICTs in practically all the spheres of knowledge, the creation of new tools to process vast amounts of data, and speed that can be generated with the dissemination and feedback, peer review peeking out from the traditional to ensure content quality.

Thus, science (and its various forms of communication) witnesses a possible transition in which higher education institutions and academic research centers must respond favorably. Given this situation, social, economic and cultural factors define various rhythms that affect scientific community groups, authorities and experts in ICT, thus creating a particular dynamic of progress. However, the review and systematization of experience could be a good contribution to stimulate collective learning and to follow the path of those who have been successful in science communication covered by ICT.

Understanding communication as an important component to strengthen the dynamics of science and to make science visible to a society to which it owes part of its funding, the idea of Open Access (OA) is gaining more and more ground. Research articles are indeed increasingly made available on a permanent and immediate form through the Web [1]. The Budapest Declaration, endorsed in 2002 by thousands of researchers, was presented with the aim of self-archiving, providing a new generation of publications to reach the idea of open access to peer-reviewed publications. One of the main arguments that support the OA initiative is that free and open access to knowledge generates in turn more knowledge and benefits to humanity, and that any control or restriction on such knowledge is an obstacle to the advancement of science [2].

One of the clearest trends that has emerged to promote this initiative are institutional repositories (IR) as a set of services committed to capture, preserve and disseminate the research conducted by knowledge communities. From the information point of view, the IR tends to be cumulative, permanent and opened, both in content and in the platform that supports it [**3**]. IR play a key role for preserving and replicating the institutional memory. This role becomes more important in the Latin-American context where most of the research group expertise stems over the figure of one or two researchers. When those experts are not present, all the expertise is lost. This fragile scientific structure forces the academic community to repeat from scratch most of the efforts to gain and to maintain the knowledge in a given area.

This paper summarizes and reports the creation of an IR at the University of The Andes (ULA), Mérida, Venezuela. Mérida is a small town (500.000 inhabitants) perched on a plateau at the Sierra Nevada, in the heart of Venezuelan Andes. Having a 200 year-old university, with 50.000 students and more than 200 research groups, it is now considered as a

"technology Mecca" in Venezuela. The geographic isolation of Mérida obliges to choose the ICT as a development tool in the mid 1980´s. Today, the United Nations Development Program reports that this small town can be considered an innovation territory where signs of technology appropriation can be detected and where ICT are statistically significant within the Latin America context [**4**].

## II. IR SABER-ULA. A SUCCESSFUL STORY

The SABER-ULA IR (2000-2008) has been developed in four stages:

### (1) Building up the infrastructure (2000-2002)

During this first phase the basic infrastructure for the repository was built. At that time we started with Alejandría software platform [5]. We also developed an aggressive information campaign about the repository services. We identified the journal editors as potential partners in the project and most of the captured contents were done replicating these university journals. The first 10 electronic journals were created in this period. Furthermore, by request of members of our university community, a service the aim of which was to publicize and disseminate the academic events of the institution was created.

### (2) Consolidating services (2002-2004)

In the second phase (2002-2004) the first services were consolidated, requirements began to emerge from the editors, and new type of services were developed depending upon the particular type of content. Some publishers began to use electronic publication as a substitute for traditional publishing, mostly due to financial and organizational problems, which delayed the publication in printed format, threatening the journal periodicity. Incentive and institutional recognition mechanisms were implemented for those producers to publish their contents in repositories, websites and open access journals in electronic format. ULA events were organized on the topic of digital libraries, and worked on the generation of models for the publication of thesis in electronic format. Some services with low acceptance at the beginning, like the researchers and research unit databases, which are part of the IR of the ULA, began to be used by various departments of the University for academic and administrative purposes. Functional tools for managing the repository were incorporated in order to ensure system interoperability with other service providers, and preparing to create content networks with other institutions.

### (3) Users and institutional recognition (2004-2006)

Between 2004 and 2006, the third stage, constant content processing volume (journal articles, pre-prints, references to events, etc.) was reached. During the first quarter of 2004 only, an average of 500 documents were processed monthly. It published the journal number 40, and 8000 information items were published in the IR. Over 60% of the content comes from the academic journals published in the repository. Users and the institution began to value the information collected in the SABER-ULA repository. For example, some ULA historians made use of the event database service to construct a memory of conferences and events held at the University. The ULA reached a relevant place regarding the visibility of its content on the Internet, largely due to the quantity and quality of its IR contents (visit the Web Ranking of World Universities and check the place occupied by the ULA among Latin American universities). In spite of the fact that full institutional recognition has not been achieved yet, an encouraging sign at the end of the first quarter of 2006 is that the ULA officially declared its commitment to sign and adhere to the Berlin Declaration, a breakthrough in understanding the importance of the ideas propagated by the OA movement and associated initiatives.

### (4) Enhancement of services (2006-2009)

In the fourth phase (2006-2008), the update of the technological platform that supports the IR of the ULA was performed. It consisted in the migration of data and services to DSpace platform developed by MIT and HP in the USA. This meant the deployment of applications that allow self-archiving as refereed publication of different types of documents and digital objects. The practice of self-archiving has not been adopted by the users' community, as originally intended, due to the usability problems of the tool, which requires adapting to local needs. In 2006 the ULA signed the Berlin Declaration. In early 2008 policies that promoted free dissemination of the intellectual production through the use of their repositories are adopted within the University. By the end of 2008, the RI SABER-ULA had recorded 15,000 documents and 60 journals. In addition, it reached 40,000,000 visits to its portal contents. A server specialized in conference management, making use of the "CDS Indico" system developed by CERN, was created. In the latest Ranking Web of World Institutional Repositories, developed by the CSIC of Spain (July 2009), the RI SABER-ULA garnered the position number 42 in the world, being the first repository of Latin America in this ranking.

Figure 1 shows the growth of the quantity of documents published at RI SABER-ULA between 2000 and 2009. It is important to note that until now (October, 2009) more than 19,000 documents and items have been stored in the repository. This graph was generated automatically by the Registry of Open Access Repositories (ROAR).

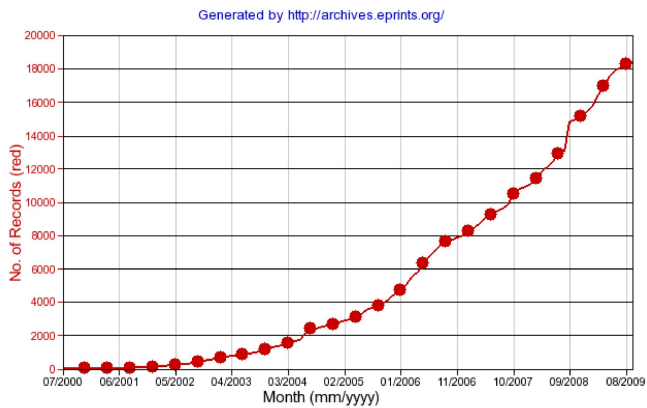

Fig. 1. Number of documents accumulated in " RI SABER-ULA", 2000-2009

Currently, at the end of 2009, we are working to facilitate the use and adoption of publishing tools and content management together with new tools that are being incorporated into the process. In the short term it will allow increasing the practice of document self-archiving. It will also permit the university community-- which generates information-- to have a better control over the content to publish and to play a more proactive role in this process.

## III. A POSSIBLE RECIPE FOR THE DEVELOPMENT OF AN INSTITUTIONAL REPOSITORY

With the idea of systematizing, organizing and transferring a decade-long experience in this area, we will now present and explain some simple strategies to be developed in order to build the info-infrastructure that could be applied in similar contexts. We thus describe the processes we followed to develop the platform and to reach the social appropriation of the tools and services of an RI. Some of these results have been reported elsewhere [6-8]. We will specifically deal with the roles established in the university organization, the various ways of conceiving scientific communication, and the institutional link with the above mentioned tools.

Although any strategy to develop an IR should be intimately adapted to the reality of the organization, we will present a possible recipe to develop a successful one. This is based on our experience and considering the actual availability of technological tools, but it has to be adapted, taking into account, not only the change in technology issues, but cultural and organizational particularities as well [9-11].

We believe that the process of appropriation of ideas and tools that promote the free dissemination of knowledge produced in our institutions can be related to the following "ingredients":

1. a Content Task Force (CTF) to create, maintain and develop a sustainable infrastructure to preserve, handle and disseminate the information
2. a sustainable infrastructure (technical & informational) to preserve, handle and disseminate the information
3. individuals and/or communities that produce information that promote the process of publication, dissemination and preservation of digital content.
4. an appropriate methodology for training these communities
5. institutional policies for handling information and incentives for the content producers

The CTF should be in charge of providing the technological ground to implement the IR. It could start as a two-specialist unit: one technical (system administrator) and the other one, a content administrator. The system administration should take care of the platform and of the server communication environment. The content admin have to deal with the service concept, the architecture of the information and the training services needed by the IR.

There are three minimum functions that the CTF should accomplish:

1. To select, adapt and test the available techniques and tools for content preservation/dissemination. The platform, techniques and tools have to be chosen and configured according to the resources, funding, technical skills, the type of content to be preserved and the service demands of the community.
2. To plan and organize the information services to be provided. In order to have a service plan, the CTF should develop a content recruitment strategy finding partners into the early adopters communities (Journal Editors, IT skillful research groups and researchers, high tech institutes or those who have their own network of international collaboration). The idea is to build a network of IR collaborators that recommend the use of its services to other colleagues [12]. It is also very important to show to the Journal Editors how the visibility through an OA repository could increase their journal impact factor [13]. Other possible strategies could be to develop new models of publishing adapted to the type of contents more commonly produced by your organization, together with the appropriate training of the users' communities.
3. To promote, publicize and to disseminate those services into the academic community. This function (and in some sense the previous one as well) should be supported by the institutional policies, mechanisms and incentives to induce the authors and contents producers to preserve/disseminate their production through the IR. It is also very important to subscribe the institutions to the OA declarations and register the IR to the corresponding directory of repositories, the search engines and integrated OA services. The idea is to make the IR be as visible as possible for all the OA community.

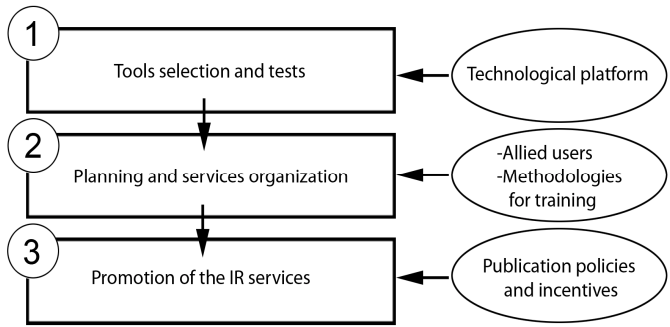

Fig. 2. Ingredients and process for developing an Institutional Repository

## IV. CONCLUSIONS

We will now present some conclusions drawn from our experience.

It is very important to select the most adequate IR platform for each institution. The pros and cons should be carefully evaluated:

- centralized vs distributed platforms
- a unique tool vs an integrated set of them
- new interoperability tools, its standards and trends

The evolution of the processes and the new expression of scholarly publishing should be studied from the beginning, in particular the impacts and incentives of self archiving on the researchers' academic promotion. In order to gain sustainability and scalability, contents managing environments (such as Open Journal System) should also be considered from the very beginning. These systems give significant advantage in promoting new services for the users.

In spite of the slow technological appropriation process for the information products, we believe that these eight years of operation of our IR have had a significant impact to start shifting the scholarly publishing model of the ULA research community. Today, this community is starting to understand the advantages of disseminating their work through OA mechanisms. We think there are some signs that reinforce this perception. The growing demand of the authors for new and better information services, the increasing consulting services and the greater exchange of information among the technical people from several technical universities are clear indications of the appropriations. It is worth mentioning that there are few IR in Latin America: the ROAR directory report only 7 IRs in Venezuela, all having not more than 90,000 documents available.

In Figure 3 we can appreciate the very low number of IR registered in LA. This can be contrasted with the corresponding number of IRs in Europe, USA and Canada. The different colors represent each platform operating the IR: Blue for Dspace (443), yellow for E-prints (317) and others for Bepress, ETD-db, Opus and Fedora.

Barriers still exist for the OA scholar publishing model. Here are some of them: few incentives for electronic publication, a very low commitment to preserve the institutional contents in electronic format, the need of certification standards for contents on the Internet, lack of skills and expertise for the appropriate use of Internet tools and publishing environments, deficient institutional policy support to promote free access to knowledge. We believe that successful OA policies and the institutional commitment with all these initiatives will help in the near future to break some of these barriers so as to foster the impact of the IR.

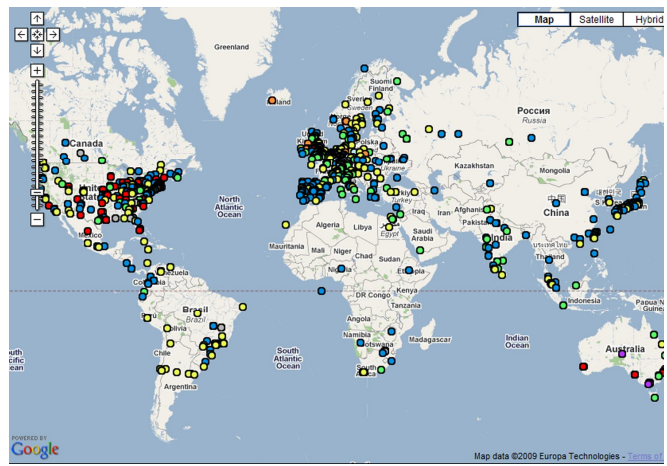

Fig. 3. Map of repositories location around the world (www.repository66.org)

Despite the success of ULA-IR and considering the discreet volume of published information, we can assert that technological infrastructure is not the main issue for implementing an IR. The organization, the publisher incentives, the content recruitment strategies and the OA promoting policies are the most important factors surrounding IR operation.

## REFERENCES

[1] E. Canessa and M. Zennaro. Science dissemination using open access. In E. Canessa and M. Zennaro, editors, *Science Dissemination using Open Access*, Trieste, Italy, 2008. International Centre for Theoretical Physics.

[2] J. Guédon. Open access archives: from scientific plutocracy to the republic of science. *IFLA journal*, Jan 2003.

[3] R. Crow. The case for institutional repositories: a sparc position paper. *ARL Bimonthly Report*, Jan 2002.

[4] PNUD. Informe sobre desarrollo humano en venezuela 2002: Las tecnologías de ia información y comunicación al servicio del desarrollo. *Informe sobre el Desarrollo Humano en Venezuela*, page 249, Aug 2002.

[5] J. G. Silva. Alejandría: De los sistemas bibliotecarios a las redes de sistemas bibliotecarios. http://biblioteca.ucv.cl/novedades/conferencias/venezuela2004/ponencias/JoseSilva.pdf, May 2004.

[6] J. A. Dávila, L.A. Núñez, B. Sandia, and R. Torrens. Los repositorios institucionales y la preservación del patrimonio intelectual académico. *Interciencia*, 31(1):22–29, 2006.

[7] J. A. Dávila, L.A. Núñez, B. Sandia, J. G. Silva, and R. Torrens. www.saber.ula.ve: un ejemplo de repositorio institucional universitario. *Interciencia*, 31(1):29–37, 2006.

[8] R. Torrens, R. Urribarri, and L.A. Núñez. ICT, community memory and technological appropriation. In Larry Stillman Graeme Johanson, editor, *Constructing and Sharing Memory: Community Informatics, Identity and Empowerment: Selected papers from the 3rd Prato International Community Informatics Conference*, pages 292–306, Newcastle, UK, 2007. Cambridge Scholars Publishing.

[9] N. F. Foster and S. Gibbons. Understanding faculty to improve ir content recruitment. *D-Lib Magazine*, 11(1), January 2005.

[10] MR Barton and MM Waters. Creating an institutional repository: Leadirs workbook. Technical report, MIT Library, 2005.

[11] K. Markey, S Rieh, B St Jean, J Kim, and E Yakel. Census of institutional repositories in the united states: Miracle project research. *Council on Library and Information Resources*, Jan 2007.

[12] A. Zuccala and C. Oppenheim. Managing and evaluating digital repositories. *Information Research*, 13(1), 2008.

[13] P. Rivas and R. Torréns. Visitas y consultas en línea a educere en su sitio web. *Educere La Revista Venezolana de Educación.*, 10(35):585–592, 2006.